\documentclass[twocolumn]{aastex631}
\usepackage{CJKutf8}
\usepackage{amsmath}	

\newcommand{\reOne}{}

\begin{document}

\title{An Eccentric Binary with a Misaligned Circumbinary Disk}

\author[0009-0000-6461-5256]{Zhecheng Hu (\begin{CJK*}{UTF8}{gbsn}胡哲程\end{CJK*})}
\affiliation{Department of Astronomy, Tsinghua University, Beijing 10084, China}

\author[0000-0003-4027-4711]{Wei Zhu (\begin{CJK*}{UTF8}{gbsn}祝伟\end{CJK*})}
\affiliation{Department of Astronomy, Tsinghua University, Beijing 10084, China}

\author[0000-0002-8958-0683]{Fei Dai (\begin{CJK*}{UTF8}{gbsn}戴飞\end{CJK*})}
\affiliation{Institute for Astronomy, University of Hawai‘i, 2680 Woodlawn Drive, Honolulu, HI 96822, USA}
\affiliation{Division of Geological and Planetary Sciences, 1200 E California Blvd, Pasadena, CA, 91125, USA}
\affiliation{Department of Astronomy, California Institute of Technology, Pasadena, CA 91125, USA}

\author[0000-0003-0853-6427]{Ping Chen (\begin{CJK*}{UTF8}{gbsn}陈平\end{CJK*})}
\affiliation{Department of Particle Physics and Astrophysics, Weizmann Institute of Science, Rehovot 7610001, Israel}

\author[0000-0003-3250-2876]{Yang Huang (\begin{CJK*}{UTF8}{gbsn}黄样\end{CJK*})}
\affiliation{School of Astronomy and Space Science, University of Chinese Academy of Sciences, Beijing 100049, People's Republic of China}

\author[0000-0001-8060-1321]{Min Fang (\begin{CJK*}{UTF8}{gbsn}房敏\end{CJK*})}
\affiliation{Purple Mountain Observatory, Chinese Academy of Sciences, 10 Yuanhua Road, Nanjing 210023, People's Republic of China}
\affiliation{University of Science and Technology of China, Hefei 230026, People's Republic of China}

\author{Richard~S.~Post} 
\affiliation{Post Observatory, Lexington, MA, USA}




\begin{abstract}

We present spectroscopic and photometric observations of Bernhard-2, which was previously identified as a candidate system to host a misaligned circumbinary disk. Our spectroscopic measurements confirm that Bernhard-2 indeed contains an eccentric ($e=0.69 \pm 0.08$) binary and thus that the periodic variability in the photometric light curve is best explained by the occultation by the misaligned circumbinary disk. \reOne{By modeling the spectral energy distributions at different phases, we infer the masses of the two binary components to be $\sim 1.1\,M_\odot$ and $\sim 0.9\,M_\odot$, respectively. The system age is determined to be $\lesssim$ 20 Myr by combining the stellar isochrone model with lithium abundance.}
Our new photometric observations show clear deviations from the model prediction based on the archival data, suggesting ongoing precession of the circumbinary disk. The H$\alpha$ line of Bernhard-2 also shows an inverse P-Cygni profile at epochs close to the pericenter passage, which could be attributed to the pulsed accretion around the pericenter.
Bernhard-2 therefore closely resembles the well studied KH 15D system. Further detailed observations and studies of such rare systems can provide useful information about disk physics and evolution.

\end{abstract}



\section{Introduction} \label{sec:intro}

\begin{figure*}[ht!]
\centering
\includegraphics[width=0.7\textwidth]{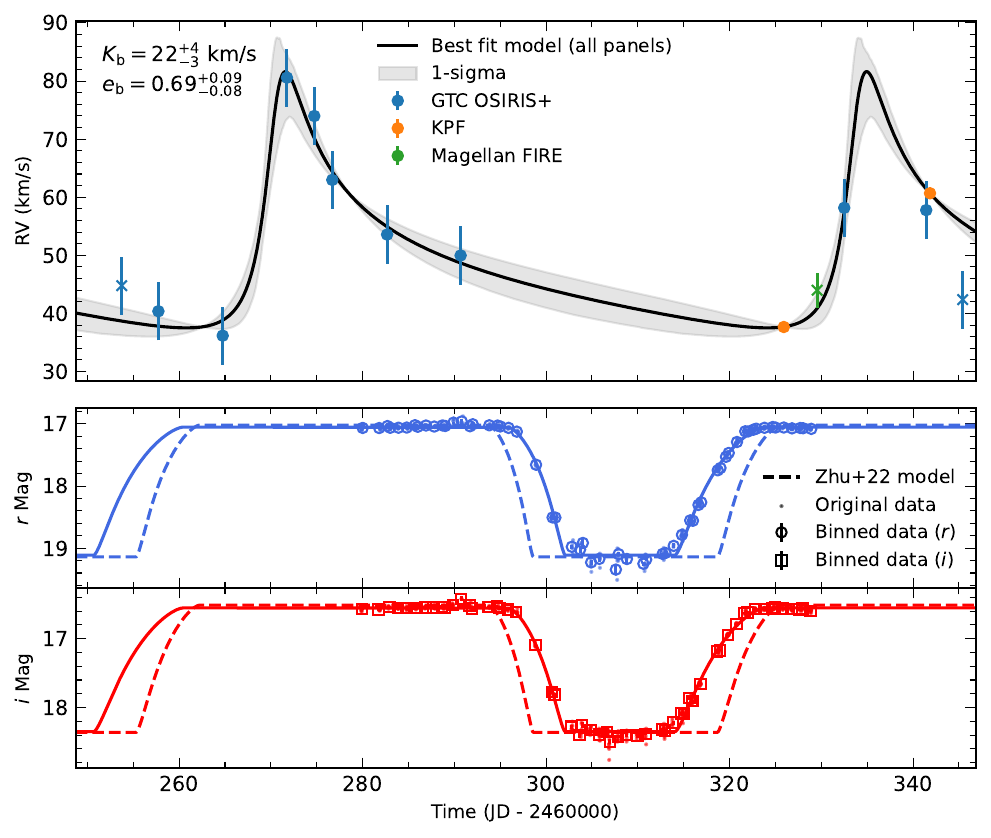}
\caption{
Spectroscopic (top panel) and photometric (middle and bottom panels) measurements of the Bernhard-2 system. The solid line in each panel represents the best-fit model. The grey shaded region in the top panel is the 1-$\sigma$ uncertainty of the RV model. In the top panel, the blue, orange, and green points are the RV measurements from GTC, KPF, and FIRE, respectively. Note that we assume the FIRE data point has the same RV offset with KPF. GTC and FIRE measurements that are not used in the RV modeling are indicated by cross. The new photometric measurements are taken in $r$ (middle panel) and $i$ (bottom panel) bands. The new light curves do not agree with the best-fit model of \citet{Zhu2022_Two}, which may suggest some substantial precession of the circumbinary disk. 
\label{fig:rv-lc}
}
\end{figure*}

The discovery of over a dozen circumbinary planets has substantially advanced our understanding of the exoplanet population \citep[e.g., ][]{Doyle2011_Kepler1st, Kostov2020_TESS1st}. Although the majority of the circumbinary planets are coplanar to the central binaries, circumbinary objects with misaligned or even polar orbits are predicted to exist, especially around eccentric binaries \citep[e.g., ][]{Aly2015_misalign, Martin2017_Polar, Lubow2018_Linear, Zanazzi2018_Inclination, Cuello2019_Planet, Smallwood2020_formation}. Indeed, some misaligned and even polar circumbinary protoplanetary disks have been detected \citep[e.g., ][]{Kohler2011_GGTauA, Kennedy2012_99Her, Brinch2016_IRS43, Lacour2016_HD142527, Fernandez-Lopez2017_SR24, Kennedy2019_HD98800, Kenworthy2022_V773TauB}, and all exhibit relatively high eccentricity ($e \gtrsim 0.5$). \reOne{Misaligned circumbinary disks are especially popular around central binaries with intermediate orbital periods (30--$10^5$ days, \citealt{Czekala2019_Degree}).}

At intermediate ages, debris disks connect the protoplanetary disk phase and the mature planetary system. Circumbinary debris disks with misaligned orbits relative to the central binaries are difficult to detect. Nevertheless, one well-studied system of this kind is KH 15D \citep{KH15D_discover}.
Following its discovery via photometric observations, theoretical models were developed to explain the photometric behavior \citep{Winn2004_KH15D, Chiang2004_KH15D}. These models suggest that a misaligned circumbinary disk, combined with the reflex motion of central binary stars, causes the system to show periodically the long and complex dimming events. Shortly after these models were proposed, spectroscopic observations confirmed the binary nature of KH 15D \citep{Johnson2004_KH}.

The unique geometric configuration of the KH 15D system allows for various observational methods to constrain different properties of the system. The system's light curve, under long-term photometric monitoring, shows complex behavior due to the precession of the disk \citep{John2005_history, Maffei2005_history, Hamilton2005_Disappearing, Capelo2012_locating, Aronow2018_Optical_Radio, GarciaSotok2020_Clumps}. This is useful for probing the disk in detail, including geometry and dynamical constraints \citep[e.g., ][]{Winn2006_Orbit, Poon2021_Constraining} and properties of disk material and substructures \citep[e.g., ][]{Silvia2008_forward_scatter, Arulanantham2016_SEEING, GarciaSotok2020_Clumps}.

The (nearly) all-sky photometric surveys have enabled the identification of more systems similar to KH 15D. In particular,
\citet{Zhu2022_Two} reported two candidate systems with KH 15D-like light curves from the Zwicky Transient Facility \citep[ZTF, ][]{Bellm2019_ZTF, Masci2019_ZTF}, which they named Bernhard-1 and Bernhard-2. Both objects exhibit dimming of more than 2 magnitudes in multiple bands, lasting over one-third of their periods. These features resemble closely the key characteristics of the KH 15D system. Additionally, both targets show infrared excess in near to mid-infrared bands, suggesting the presence of a cold disk. 

In this work, we confirm with spectroscopic observations that the Bernhard-2 system indeed contains an eccentric binary. In Section \ref{sec:obs}, we describe our new observations and the data reduction procedures. In Section \ref{sec:physical}, we present the physical interpretations, including modeling of the RV curve and the system SED. Finally, in Section \ref{sec:r-and-d}, we summarize our results and discuss a few interesting features that deserve future follow-up studies.

\section{Observations}
\label{sec:obs}

\subsection{GTC/OSIRIS}\label{subsec:gtc-obs}

We obtained 11 spectra of Bernhard-2 over a time interval of $\sim 92$ days from the OSIRIS instrument \citep{Cepa2013_Highlights} installed on the 10.4\,m Gran Telescopio CANARIAS (GTC). Observations were taken at random orbital phases during the out-of-occultation windows. A radial velocity standard star, CoRot-7, was also observed following Bernhard-2 in the first three nights in order to test the RV stability. 
We used the long-slit mode of OSIRIS, with the R2500R grism and a slit width of 0.6$\arcsec$ for both targets. This yields a wavelength coverage from 5575 ${\rm \AA}$ to 7685 ${\rm \AA}$ and a spectral resolution of $\sim 2500$. The exposure times are 900\,s and 20\,s for Bernhard-2 and CoRot-7, respectively. 

The GTC data were reduced with the \texttt{PypeIt}\footnote{\url{https://pypeit.readthedocs.io/en/latest/}} package \citep{pypeit:joss_pub,pypeit:zenodo} following the standard process. The wavelengths are calibrated by the lamp spectra taken by the same instrument. For each lamp frame, we manually identified strong lines and ignored those with signal-to-noise (S/N) below 20, in order to achieve a robust wavelength solution. 

The RV standard star shows an RV variation up to 20\,km/s after the wavelength calibration by the lamp spectra, so we also refined the wavelength solution by calibrating the telluric emission lines to the sky spectrum model. In this step, the telluric lines were obtained by \texttt{PypeIt}, and the wavelengths of the sky spectrum model, available on the GTC website
\footnote{\url{https://www.gtc.iac.es/instruments/osiris+/media/sky/sky_res2500.txt}}
were converted from air to vacuum by the \texttt{airvacuumvald}\footnote{\url{https://pypi.org/project/airvacuumvald/}} package, in which step the relation in \citet{Morton2000_Atomic} was used. After this refinement, the RV uncertainty on CoRot-7 was reduced substantially, with a standard deviation of $\sim$3.5\,km/s from the absolute RV value of the same star from \emph{Gaia} DR3 \citep{Gaia2023_DR3}. Applying the same procedure to Bernhard-2, we obtained a wavelength uncertainty of $\sim$0.1\,$\rm \AA$ based on the internal scattering.

We extract the RVs by cross-correlating the wavelength-calibrated stellar spectra with a single-star spectrum template. The template is obtained by interpolating in the \texttt{Phoenix} library \citep{Husser2013_PHOENIX} at the parameter values of the primary star. Motivated by the original spectral analysis of \citet{Zhu2022_Two} and the SED fitting in Section \ref{subsec:sed}, we set the primary star parameters to be $T_{\rm eff} = 4850\,$K, $\log g = 4.5$, and $\rm [Fe / H]=0.4$. The theoretical template is further \reOne{convolved} to the resolution of ${\rm R} \sim 2500$. Before conducting the cross-correlation, we normalize both the observed and template spectra using \texttt{iSpec} \citep{Blanco2019_iSpec}, and visually confirm that the resulting continua are flat. Additionally, we have removed outliers and excluded certain wavelength ranges that were affected by the atmosphere (e.g., oxygen $\gamma$ band) or the stellar variability (e.g., H$\alpha$; see Section \ref{sec:r-and-d}). In deriving the RV values we have not considered the impact of the uncertainties in the primary star parameters or the inclusion of the secondary star. The former would only modify the line profiles slightly, and its impact on the derived RVs is negligible. Regarding the impact of the secondary star, this secondary star only contributes $\sim 0.16$ to the total flux in the wavelength range that is considered here (see Section~\ref{subsec:sed}). We have tested that using a binary spectrum template with such a flux ratio would only change the derived RV values typically by $\lesssim 3\,$km/s, which is comparable to the RV uncertainty from the wavelength calibration. The contribution of photon noise to the final RV error is determined through bootstrapping, yielding an error of $\lesssim 1.5\,$km/s across all epochs. The overall RV uncertainty is set at $5\,$km/s, significantly exceeding the photon-noise-limited error to accommodate potential systematics. The extracted RV measurements are shown in the top panel of Figure \ref{fig:rv-lc}, and the values are also given in Table \ref{tab:rv-data}.

\subsection{Other Spectroscopic Measurements}\label{subsec:kpf-obs}

Two spectroscopic observations of Bernhard-2 were obtained from the newly commissioned Keck Planet Finder \citep[KPF, ][]{Gibson2016, Gibson2018, Gibson2020} mounted on the Keck I telescope. KPF is a fiber-fed echelle spectrograph with a spectral resolution of $R \sim 98,000$ and a wavelength coverage from 4450 ${\rm \AA}$ to 8700 ${\rm \AA}$. Both observations had an exposure time of 900 seconds. The spectra were reduced with the KPF Data Reduction Pipeline (DRP) that is publicly available \footnote{\url{https://github.com/Keck-DataReductionPipelines/KPF-Pipeline}}. See also the RV measurement  by \citet{Dai2024_KPF}. Because the KPF pipeline estimates RV error based solely on the photon noise, it tends to underestimate the errors for faint stars such as Bernhard-2. To address this issue, we choose to inflate the RV uncertainties of KPF by approximately a factor of five. Additionally, we have also incorporated an RV jitter term in the modeling, as detailed in Section \ref{subsec:rv-model}. 

We have also obtained one spectrum of Bernhard-2 in the near IR with Magellan Folded-port InfraRed Echellette (FIRE; \citealt{Simcoe2008SPIE}) using the echelle mode, which delivers continuous wavelength coverage from $Y$ through $K$ bands (0.82-2.51 microns). The observation was conducted with 1$\arcsec$ width slit under varying seeing between 1.0$\arcsec$ to 1.4$\arcsec$. The FIRE data was reduced with the \texttt{PypeIt} package following the standard procedures. The OH sky emission lines were used for wavelength calibration. The estimated resolution of the result spectrum is $\sim$2500. Using a stellar template with $T_{\rm eff}=4900\,$K, we derive an RV measurement of $44\pm3\,{\rm km\,s^{-1}}$.

These KPF and FIRE RV measurements are also shown in the top panel of Figure~\ref{fig:rv-lc}.

\subsection{Photometric Observations}\label{subsec:phot-obs}

High-cadence photometric observations of Bernhard-2 were obtained between December 1st, 2023, and January 19th, 2024, when the occultation event was expected to happen. \reOne{These observations were taken in both $r$ and $i$ bands by the 32 inch telescope at the Post Observatory in Lexington, Massachusetts, USA}. As shown in the middle and bottom panels of Figure~\ref{fig:rv-lc}, these new light curves appear different from the prediction of the original photometric model of \citet{Zhu2022_Two}, which was based on the sparse ZTF observations, again suggesting that the misaligned circumbinary disk may be undergoing precession. The updated times and projected velocities of both ingress and egress of the occultation event are given in Table~\ref{tab:star-prop}. Intensive photometric observations like those shown in Figure~\ref{fig:rv-lc} in the long term will be useful in better constraining the time evolution of such a rare system, as has been demonstrated in the famous KH-15D system (see \citealt{Poon2021_Constraining} and references therein). \reOne{They can also be used to study the opaqueness of the circumbinary disk, as will be further discussed in Section~\ref{sec:lc}.}

\section{Physical Interpretation} \label{sec:physical}

\begin{table}
\centering
\caption{The SED and RV best-fit parameters as well as the inferred properties (marked with $^*$). The subscript $1$ and $2$ indicate the primary and secondary star in the binary, respectively. Here BJD$^\prime=$BJD$-2460000$.}
{\footnotesize
\begin{tabular}{lll}
\hline\hline
\multicolumn{3}{l}{\bf{SED fitting}}   \\ 
S1 EEP &                    EEP$_{1}$ &       $183 \pm 3$ \\
S2 EEP &                    EEP$_{2}$ &       $173 \pm 4$ \\
Log of age (yr) &          $\log \mathrm{Age}$ &    $7.3 \pm 0.09$ \\
Metallicity &            $\mathrm{[Fe/H]}$ &   $0.46 \pm 0.11$ \\
Distance &                    $d$ (kpc) &   $1.82 \pm 0.13$ \\
Extinction &                  $A_V$ (mag) &    $0.85 \pm 0.10$ \\
S1 Mass &    $M_{1}^{*}$ ($M_{\odot}$) &   $1.13 \pm 0.07$ \\
S1 Radius &    $R_{1}^{*}$ ($R_{\odot}$) &   $1.15 \pm 0.08$ \\
S1 Effective Temperature & $T_{\mathrm{eff},1}^{*}$ (K) &     $4870 \pm 70$ \\
S2 mass &    $M_{2}^{*}$ ($M_{\odot}$) &   $0.87 \pm 0.04$ \\
S2 radius &    $R_{2}^{*}$ ($R_{\odot}$) &   $1.15 \pm 0.08$ \\
S2 Effective Temperature & $T_{\mathrm{eff},2}^{*}$ (K) &     $3990 \pm 40$ \\
\hline
\multicolumn{3}{l}{\bf{RV fitting}}   \\ 
         Periastron time &       $T_{P1}$ (BJD$^\prime$) & $270.5 \pm 0.8$ \\
            Eccentricity &                          $e$ & $0.69 \pm 0.08$ \\
  Argument of periastron &           $\omega_{1}$ (deg) & $316 \pm 10$ \\
       RV semi-amplitude &               $K_{1}$ (km/s) &      $22 \pm 4$ \\
   System velocity (GTC) &    $\gamma_{\rm GTC}$ (km/s) &  $53.8 \pm 2.3$ \\
   System velocity (KPF) &    $\gamma_{\rm KPF}$ (km/s) &  $48.4 \pm 1.5$ \\
 Inclination &                 $i^{*}$ & $36 \pm 5$ \\
\hline
\multicolumn{3}{l}{\bf{Light Curve fitting}}   \\ 
Ingress Time & $t_{\mathrm{in}}$ (MJD) & $59158.04 \pm 0.03$ \\
Egress Time & $t_{\mathrm{out}}$ (MJD) & $59177.97 \pm 0.02$ \\
Projected Ingress Velocity & $v_{\mathrm{in}}$ ($R_{\star}/$day) & $0.323 \pm 0.005$ \\
Projected Egress Velocity & $v_{\mathrm{out}}$ ($R_{\star}/$day) & $0.2065 \pm 0.0015$ \\
\hline
\hline
\end{tabular}
}
\begin{flushleft}
\label{tab:star-prop}
\end{flushleft}
\end{table}

\begin{figure*}[ht!]
\includegraphics[width=\textwidth]{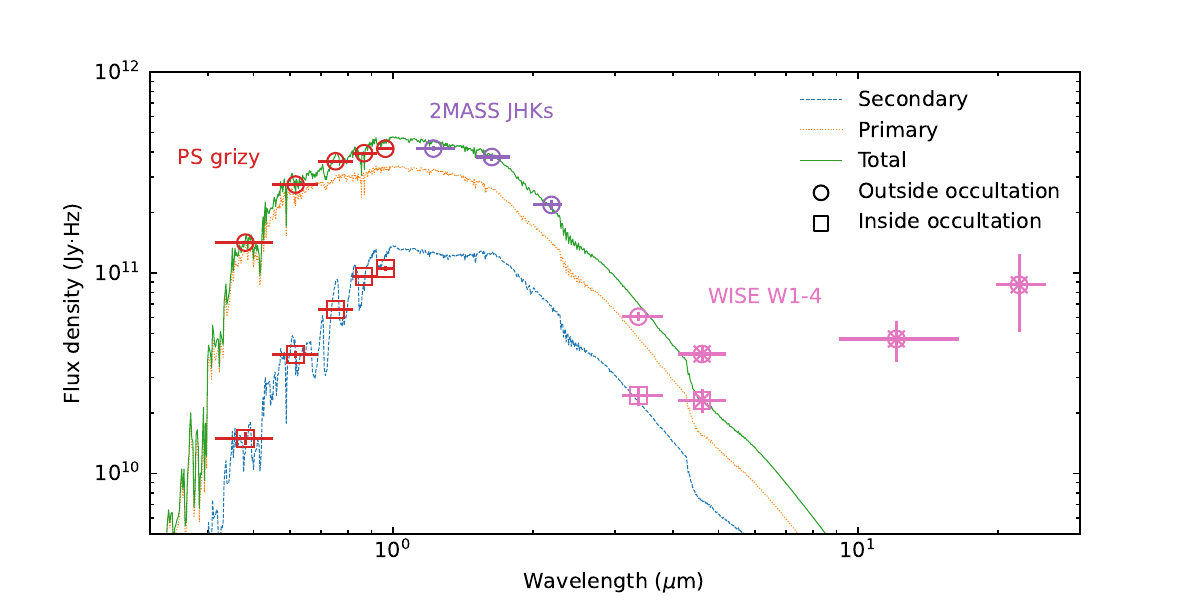}
\caption{
SEDs at different phases and the best-fit binary model. Observations from Pan-STARRS, 2MASS, and WISE are shown in red, purple, and pink, respectively. The extent of each point along the wavelength direction indicates the corresponding bandwidth. The data obtained outside-occultation phase is shown in circles, while the data obtained inside-occultation phase is shown in squares. Note that points with an additional cross are not used in the fitting process.
The orange dotted and blue dashed and lines show the best-fit stellar SED model for the primary and secondary, respectively. The outside-occultation data are fitted by the summation of the two components, which is shown in green solid line.
\label{fig:SED}
}
\end{figure*}

\subsection{Radial Velocity Modeling}
\label{subsec:rv-model}

We model the radial velocity data with a Keplerian model using the \texttt{radvel} package \citep{Fulton2018_RadVel}. \reOne{The RV model is described by six parameters, including three that determine the shape of the curve---namely the RV semi-amplitude $K$, the argument of periastron $\omega$, and the orbital eccentricity$e$---and another three that determine the vertical and horizontal offsets of the model, i.e., the time of pericenter passage $T_{\rm P}$, the velocity offset for GTC $\gamma_{\rm GTC}$, and the velocity offset for KPF $\gamma_{\rm KPF}$.}

\begin{figure*}[ht!]
\centering
\includegraphics[width=0.7\textwidth]{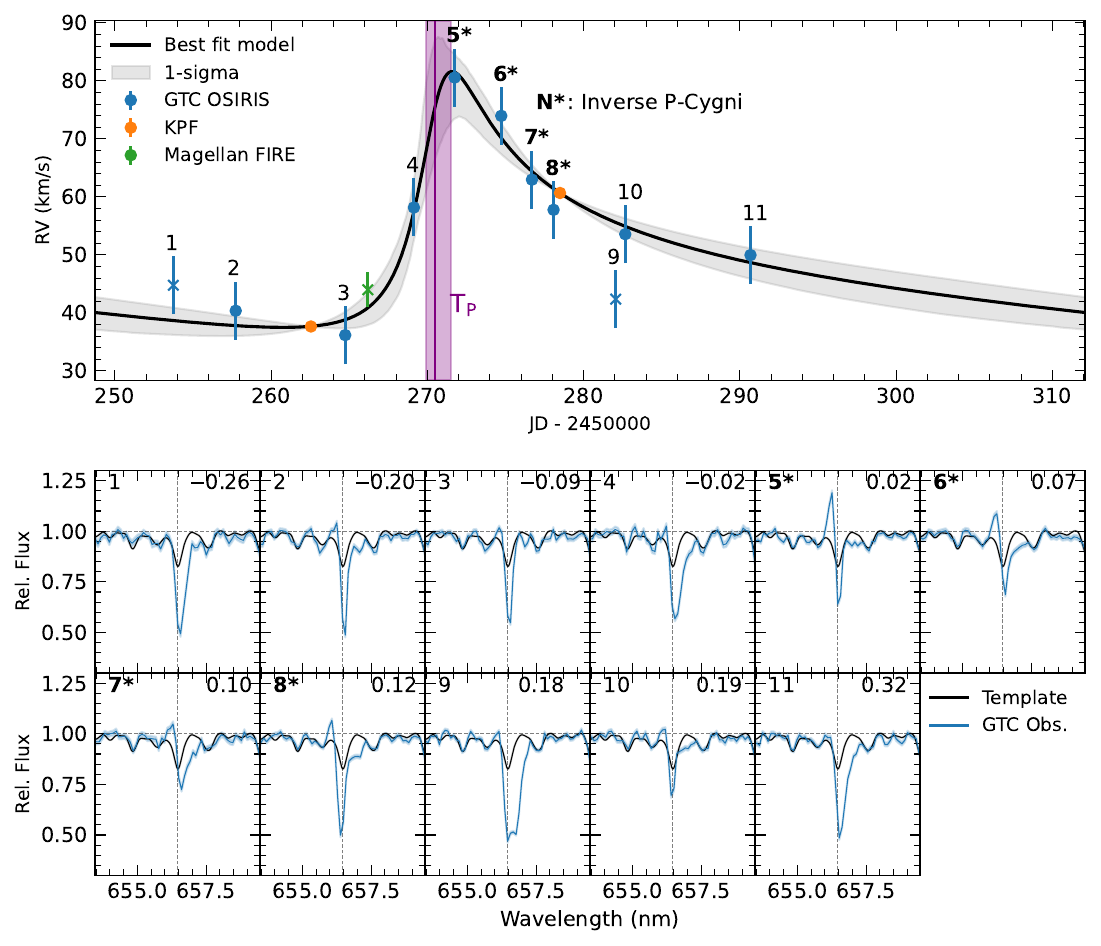}
\caption{
The phase folded RV fitting result (upper panel) and the corresponding GTC H-$\alpha$ profiles (lower panel). The description of data points and model in the upper panel is the same as Figure \ref{fig:rv-lc} except for the number above each GTC data point, which labels the observations in ascending phase order. Note that the bold numbers with asterisk are the observations with inverse P-Cygni H-$\alpha$ profile. The pericenter passage time is shown with the purple solid line, respectively. The purple shaded region indicates the fitting error of ${\rm T}_{\rm P}$. In the lower panels, we show the observed spectra as well as the single star template near the H$\alpha$ region with blue and black solid lines, respectively. The number on the upper left of each panel is the observation label in the upper panel, while the number on the upper right indicates the corresponding phase.
\label{fig:rv-halpha}
}
\end{figure*}

We have fixed the orbital period to 63.358 days, which is the best-fit value in \citet{Zhu2022_Two}, as we do not expect to constrain $P$ from the sparse RV observations alone. We have nevertheless tested that the derived model parameters are not affected by any small change in the period value we use. The RV jitters of GTC and KPF observations are consistent with zero if included, so for simplicity we do not include the jitter term. We fit the systemic velocities for GTC and KPF data separately to compensate for the velocity offset between the two telescopes. Since there is only one data point from Magellan/FIRE, it is used for a sanity check and not included in the fitting. It aligns with the model prediction to within 1-$\sigma$, irrespective of whether we assume it shares the same velocity offset with GTC or KPF.
As shown in the top panel of Figure~\ref{fig:rv-lc}, the first and last GTC observations are not included in the RV modeling. The former was observed during egress and the latter is a clear outlier. Therefore, we have in total six free parameters for 11 data points. \reOne{We confirm that the model parameters, especially $K$ and $e$, are hardly changed even if the outlier data point is included.}
We use \texttt{emcee} \citep{Foreman-Mackey2013_emcee} to sample the posteriors of the model parameters. In total 100 walkers are used, with each drawing 7000 steps and the first 5000 steps removed as burn-in steps. The best-fit parameters with the associated uncertainties are given in Table \ref{tab:star-prop}, and the best-fit model is shown in the upper panel of Figure~\ref{fig:rv-lc}. The phase-folded RV model and data are also shown in Figure~\ref{fig:rv-halpha}.

With $e=0.69\pm0.08$ and $K=22\pm4\,{\rm km~s}^{-1}$, Bernhard-2 is indeed a stellar binary with \reOne{a} very eccentric orbit, thus resembling the key characteristics of the KH-15D system. We next model the system SEDs in order to determine the physical parameters of the binary stars.

\subsection{Spectral Energy Distribution} \label{subsec:sed}

\begin{figure*}[ht!]
\centering
\includegraphics[width=0.8\textwidth]{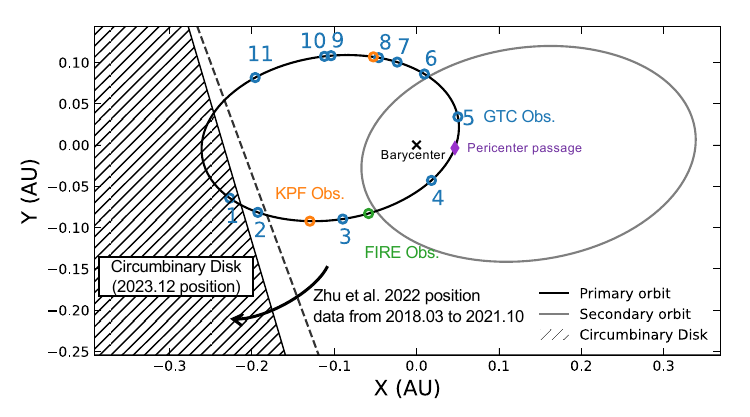}
\caption{
The sky-projected view of the binary orbital dynamics and the boundary of the occulting disk in the Bernhard-2 system. In order to illustrate the orbital characteristics and the change in the disk edge position, we only show the model with the best-fitting parameters. The hatched area shows the updated position of the disk edge from our new photometric observations (detailed in Section \ref{subsec:phot-obs}), whereas the dashed line indicates the edge position reported in \citet{Zhu2022_Two}, which utilized ZTF data collected over the period from 2018 to 2021. The black and gray ellipses represent the primary and secondary orbits, respectively, which are constructed using the best-fitting RV and stellar parameters, assuming an arbitrary longitude of the ascending node. The blue, orange, and green circles denote the positions of the primary star when the observations were conducted with GTC OSIRIS, KPF, and Magellan FIRE, respectively. Similar to Figure \ref{fig:rv-halpha}, the GTC observations are indexed according to the orbital phases.
\label{fig:show-orbit}
}
\end{figure*}

We adopt the multi-band photometric data from \citet{Zhu2022_Two} and conduct a spectral energy distribution (SED) fitting, in order to determine the physical parameters of the two stellar components. The data include \textit{grizy} bands from Pan-STARRS1 \citep{Chambers2016_PS1}, \textit{JHK} bands from 2MASS \citep{Skrutskie2006_2MASS}, and\textit{W1} to \textit{W4} bands from WISE \citep{Wright2010_WISE}, ranging from $\sim$ 0.5 $\micron$ to $\sim$ 20 $\micron$ for both inside and outside the occultation. We include a minimum photometric accuracy of $0.03$ mag to account for the potential systematics between different surveys, which is added in quadrature to the reported photometric uncertainty to derive the total photometric error (see also \citealt{El-Badry2024_M_}). 

The SED of the Bernhard-2 system shows clear evidence for the NIR excess \citep{Zhu2022_Two}, so in modeling the stellar SED, we have excluded the WISE \textit{W2} to \text{W4} measurements. 
We have tested that the derived stellar parameters and the corresponding errors remain largely unchanged if the WISE \textit{W1} band is also excluded. Under our assumption of a fully opaque disk, the inside-occultation SED is produced by the secondary star alone, whereas the outside-occultation SED is the superposition of both stellar components. For the primary star, we have also included the spectroscopic measurements from \citet{Zhu2022_Two}, namely the effective temperature $T_{\rm eff} = 4865 \pm 85$ K and the surface gravity $\log g = 4.37 \pm 0.04$, and the parallax constraint of $0.34 \pm 0.16$ mas from \textit{Gaia} DR3 \citep{Gaia2023_DR3}.

We modeled the stellar SED with the \texttt{isochrones}\footnote{\url{https://isochrones.readthedocs.io/en/latest/}} package \citep{Morton2015_isochrones}. The \texttt{isochrones} package interpolates on the MESA Isochrones \& Stellar Tracks (MIST) \citep{Dotter2016_MIST,Choi2016_MIST} to predict stellar physical parameters. 
The magnitude of a star in a certain band is given by
\begin{equation}
    m_{\rm B} = f({\rm EEP}, \log {\rm age}, {\rm [Fe/H]}, d, A_V) .
\end{equation}
Here EEP is the equivalent evolution phase \citep{Dotter2016_MIST}, $d$ is the distance, and $A_V$ is the extinction. The standard \citet{Cardelli1989_extinction} extinction law with $R_V=3.1$ is assumed. The two stars in the binary share all properties except for the EEP parameter.
The best-fit model is obtained by optimizing the total log-likelihood of all measurements, and again we use
the \texttt{emcee} sampler \citep{Foreman-Mackey2013_emcee} to sample the posterior and derive the uncertainties of the model parameters.

The result of the stellar SED fitting is also given in Table \ref{tab:star-prop}, and the corresponding stellar spectra generated with the best-fit stellar parameters are shown in Figure \ref{fig:SED} with the dashed line. For a better illustration, we have used a spectra interpolator \texttt{pystellibs}\footnote{\url{https://mfouesneau.github.io/pystellibs/}} to generate the spectra, and the empirical stellar spectra library BaSeL \citep[v2.2, ][]{Lejeune1997_Standard,Lejeune1998_Mdwarf} has been used. 

Our results indicate that the binary is composed of an early K-type star of $\sim1.1\,M_\odot$ and a late K-type star of $\sim0.9\,M_\odot$ at a distance of $\sim 1.8\,$kpc. \reOne{We note that the stellar mass derived for pre-main sequence stars could vary between different evolution tracks, and this systematic uncertainty is not accounted for here.} The stellar age is estimated to be \reOne{$20\pm4$}\,Myr, in general consistent with the young nature of the system. 
\reOne{This isochrone age is also broadly consistent with the age determined from the Li I line at 6708 $\text{\AA}$ (see Appendix~\ref{app:li}).}

The predicted flux ratio of the binary in the optical $r$ band is around 0.16, so the impact of the secondary star on the RV derivation of the primary star is limited. However, the flux ratio in the IR bands (e.g., $W1$) can be as large as $\sim 0.7$, which may be large enough to identify the secondary component directly. 

When combined with the eccentricity and RV semi-amplitude measurements from RV modeling, the derived stellar masses allow us to constrain the inclination of the binary orbit. This yields $\sin{i}=0.58 \pm 0.07$ and $i=36 \pm5 ^\circ$. Additionally, the binary separation can be derived through Kepler's third law
\begin{equation}
\begin{aligned}
a_{\rm bin} = 0.4 \ {\rm AU} \left( \frac{P_{\rm bin}}{63 \ {\rm day}} \right)^{2/3} \left( \frac{M_1 + M_2}{2 \ M_\sun} \right)^{1/3}.
\end{aligned}
\end{equation}

The near-IR excess is expected to originate from the circumbinary disk. \reOne{Detailed} modeling of the disk spectral energy distribution is not possible with the available observations, but as a start it is reasonable to believe that the radiation from the circumbinary disk peaks beyond $\sim 20\,\mu$m based on the WISE measurements. This corresponds to a blackbody temperature $\lesssim 140\,$K, which is the equilibrium temperature at a distance of $\gtrsim 4\,$au from the central binary. Therefore, the circumbinary disk in Bernhard-2 may extend out to several au, \reOne{similar} to the prototype KH 15D system \citep{Poon2021_Constraining}.
\footnote{For comparison, the KH 15D system does not have such clear near-IR excess \citep{Arulanantham2016_SEEING}.}

\subsection{Light Curve} \label{sec:lc}

\reOne{The densely sampled light curve allows us to constrain the opaqueness of the circumbinary disk, as it occults the primary star. The light curve model of \citet{Zhu2022_Two} consists of an opaque screen with a sharp edge and a binary star system behind it. The light curve is given by:
\begin{equation}
F_\lambda(t) = F_{1,\lambda} \cdot \eta(t) + F_{2,\lambda} ,
\end{equation}
where $F_{1,\lambda}$ and $F_{2,\lambda}$ are the stellar fluxes of the primary and secondary stars at a given wavelength $\lambda$, respectively. With the assumption that the secondary star is always visible and that the disk is fully opaque, the parameter $\eta(t)$, which varies between zero and one, denotes the fraction of the primary star that is not occulted and should be wavelength-independent.}

\begin{figure}[ht!]
\centering
\includegraphics[width=0.5\textwidth]{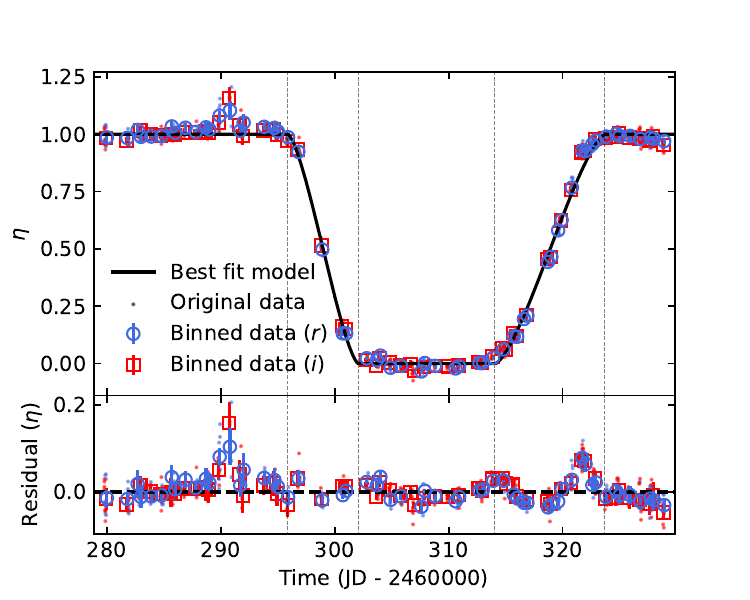}
\caption{
\reOne{The change of the fraction that the primary is not occulted, i.e., $\eta$, as a function of time. The gray dashed lines indicates the start and end of the ingress and egress, respectively.}
}
\label{fig:lc-color}
\end{figure}

\reOne{The reconstructed $\eta$ curves in the $r$ and $i$ bands are shown in Figure \ref{fig:lc-color}. The two curves are nearly identical, confirming that the occultation process is independent of wavelength and thus the disk is opaque (at least in $r$ and $i$ bands). There are some wiggles during egress that cannot be well fitted by an opaque screen with a sharp edge. This may suggest that the disk edge is clumpy, as is also observed in KH 15D \citep[e.g.,][]{GarciaSotok2020_Clumps}. More multi-band photometric observations with high S/N and cadence can help us better constrain the disk properties.}

\section{Discussion} \label{sec:r-and-d}

In this paper, we present new spectroscopic and photometric observations of the Bernhard-2 system, which has been proposed as a KH 15D-like system. The spectroscopic observations confirm that Bernhard-2 is indeed a highly eccentric binary ($e = 0.69 \pm 0.08$), and the photometric behavior can be explained by a circumbinary disk that is (potentially highly) misaligned relative to the binary orbit. Therefore, Bernhard-2 is a confirmed KH 15D-like system. The system configuration is illustrated in Figure \ref{fig:show-orbit}. Through modeling the SEDs inside and outside the disk occultation, we determine the binary to be composed of two pre-main sequence K-dwarfs of $\sim 1.1\,M_\odot$ and $\sim 0.9\,M_\odot$, respectively.

\reOne{The} spectroscopic observations deliver more than the RV measurements. As shown in the lower panels of Figure~\ref{fig:rv-halpha}, the H-$\alpha$ line profiles from GTC OSIRIS appear to be time-variable. In comparison with the model template, which is shown in black, the observed H-$\alpha$ lines are deeper and wider, and the depth and width vary with time. Similar to the KH 15D system \citep{Hamilton2012_Complex}, Bernhard-2 also shows an inverse P-Cygni profile in the H-$\alpha$ line at certain epochs. The blue-side emission component is clearly seen in epochs 5 and 6 and marginally visible in epochs 7 and 8. These epochs are close to the time of the pericenter passage, which is marked by the purple vertical line in Figure~\ref{fig:rv-halpha}, consistent with the theoretical modelings that the pulsed accretion phenomenon is enhanced during or after the pericenter passage \citep[e.g.,][]{Artymowicz:1996, Gunther:2002, deValBorro:2011}. \reOne{This picture is not changed even if another accretion tracer (e.g., He I line at 1.083\,$\mu$m; \citealt{He10831}) is used. The He I line in the Magellan/FIRE spectrum, which was taken before the pericenter passage, reveals a normal line profile with no emission component. Spectroscopic} observations with higher S/N and resolutions are needed to better understand the behavior and origin of the H-$\alpha$ line variations.

Our new photometric observations show very clear deviation, most notably in the shortened occultation duration, from the model prediction based on the ZTF data. This may be indicating the ongoing precession of the circumbinary disk, similar to that seen in the prototype KH 15D system. Long-term photometric monitoring of Bernhard-2 is needed to better understand and constrain the dynamical evolution of this rare system.

\section*{Acknowledgments}
We thank Guo Chen, Greg Herczeg, and Sharon Xuesong Wang for useful discussions.
\reOne{We thank the anonymous reviewer for comments and constructive suggestions.}
This work is supported by the National Natural Science Foundation of China (grant No. 12173021 and 12133005).
The GTC data were taken by the program GTC3-23ACNT under the agreement between GTC and the National Astronomical Observatories of China.

%

\vspace{5mm}
\facilities{GTC(OSIRIS), Keck I(KPF), Magellan(FIRE)}


\software{astropy \citep{astropy:2013, astropy:2018, astropy:2022}, emcee \citep{Foreman-Mackey2013_emcee}, isochrones \citep{Morton2015_isochrones}, pypeit \citep{pypeit:joss_pub, pypeit:zenodo}, radvel \citep{Fulton2018_RadVel}, iSpec \citep{Blanco2019_iSpec}
}



\newpage

\appendix

\section{Age determined by Lithium Abundance}
\label{app:li}

\begin{figure*}[ht!]
\centering
\includegraphics[width=0.8\textwidth]{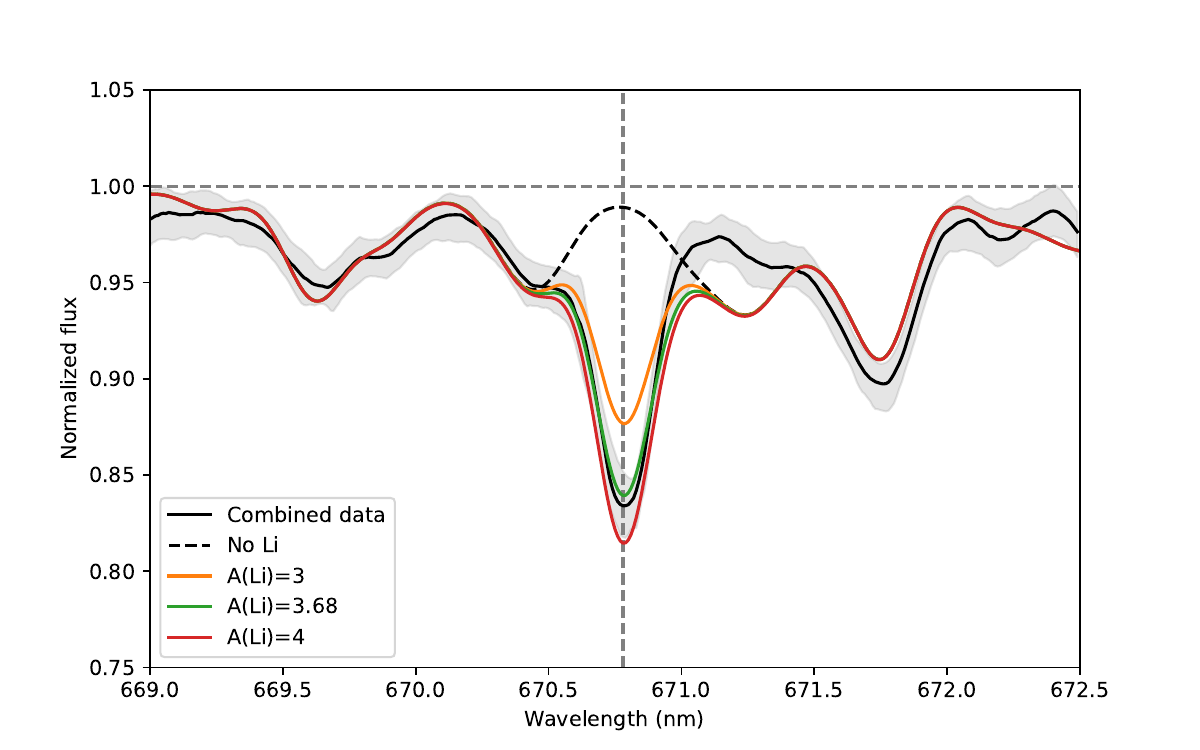}
\caption{
\reOne{The combined GTC spectrum and a few synthetic spectra around the 6708 $\text{\AA} \ \rm{Li\ I}$ line. The black solid line represents the combined data, with the gray region indicating the 1-sigma uncertainty. The black dashed line shows the spectrum with no lithium. The spectra in orange, green, and red have lithium abundances $\rm{A(Li)} = 3.0$, 3.68 (the best-fit value), and 4.0, respectively. The horizontal dashed line indicates the continuum level, while the vertical dashed line marks the center of the 6708 $\text{\AA}$ $\rm{Li\ I}$ line.}
}
\label{fig:li}
\end{figure*}

\reOne{Lithium abundance can serve as a chemical clock to verify the young nature of the system. We measure the lithium abundance and equivalent width (EW) using the 6708 $\text{\AA}$ Li I line in the combined GTC spectrum. Based on spectral synthesis using the radiative transfer code \texttt{MOOG} \citep{Sneden2012_Moog} and the Kurucz grid of ATLAS9 model atmospheres \citep{Castelli2003_ATLAS9}, we determine the Lithium abundance $\rm{A(Li)} = 3.68 \pm 0.05$. The EW is calculated from the difference between the model spectra and a lithium-depleted spectrum, yielding $418 \pm 12$ m$\text{\AA}$. The combined spectrum and the synthetic spectra with selected values of A(Li) are shown in Figure \ref{fig:li}.}

\reOne{The high A(Li) value suggests that the Bernhard-2 system is young. To quantify it, the age of the system is determined by the measured EW, based on the model and the Bayesian code from \citet{Jeffries2023_Gaia}. We adopt a log-uniform prior between $10^6$--$10^{10}$\,yr on the stellar age. Although the stellar age is not well determined, we are able to set a 95\% upper limit to 19 Myr. This is slightly younger than, although still generally consistent with, the system age derived from the SED fitting.}

\section{The RV Data}

The RV measurements are provided in Table~\ref{tab:rv-data}, in case they are needed for future analysis.

\begin{table}[ht!]
\centering
\caption{
    The extracted RV for the Bernhard-2 system.
}
\begin{tabular}{ccc}
\hline
\hline
${\rm BJD}_{\rm TDB} - 2460000$ & RV (km/s) & Instrument \\
\hline
253.7506 & $50\pm5$ & GTC OSIRIS  \\
257.7451 & $46\pm5$ & GTC OSIRIS  \\
264.7550 & $42\pm5$ & GTC OSIRIS  \\
271.7586 & $86\pm5$ & GTC OSIRIS  \\
274.7413 & $79\pm5$ & GTC OSIRIS  \\
276.7037 & $68\pm5$ & GTC OSIRIS  \\
282.6777 & $59\pm5$ & GTC OSIRIS  \\
290.6924 & $55\pm5$ & GTC OSIRIS  \\
325.9211 & $37.67\pm0.10$ & KPF  \\
329.5608 & $44\pm3$ & \reOne{Magellan} FIRE  \\
332.5107 & $64\pm5$ & GTC OSIRIS  \\
341.4479 & $63\pm5$ & GTC OSIRIS  \\
341.8599 & $60.68\pm0.10$ & KPF  \\
345.4235 & $48\pm5$ & GTC OSIRIS  \\
\hline
\end{tabular}
\label{tab:rv-data}
\end{table}


\bibliography{Bernhard_RV}{}
\bibliographystyle{aasjournal}



\end{document}